# A proposal for measuring the structure of economic ecosystems: a mathematical and complex network analysis approach

Marcelo S. Tedesco† Miguel Angel Nunez Ochoa‡ Francisco Javier Ramos Soria§

Olga Medrano Martín del Campo¶ Kevin Beuchot±

July 9, 2022

**Abstract**

The benefits of using complex network analysis (CNA) to study complex systems, such as an economy, have become increasingly evident in recent years. However, the lack of a single comparative index that encompasses the overall wellness of a structure can hinder the simultaneous analysis of multiple ecosystems. A formula to evaluate the structure of an economic ecosystem is proposed here, implementing a mathematical approach based on CNA metrics to construct a comparative measure that reflects the collaboration dynamics and its resultant structure. This measure provides the relevant actors with an enhanced sense of the social dynamics of an economic ecosystem, whether related to business, innovation, or entrepreneurship. Available graph metrics were analysed, and 14 different formulas were developed. The efficiency of these formulas was evaluated on real networks from 11 different innovation-driven entrepreneurial economic ecosystems in six countries from Latin America and Europe and on 800 random graphs simulating similarly constructed networks.

†Corresponding author: Marcelo S. Tedesco is a Research Affiliate at the MIT D-Lab of the Massachusetts Institute of Technology, where he is studying the social dynamics of economic ecosystems with the Local Innovation Group. He is the founder and Executive Director of Global Ecosystem Dynamics (GED), an international research initiative focused on advancing the study of economic ecosystems. He holds a PhD in Social & Business Science and has also studied business management, philosophy, and social anthropology. Mail: tedesco@mit.edu

‡Miguel Angel Nunez Ochoa obtained a BSc in Biomedical Engineering from the Monterrey Institute of Technology, a MicroMasters certification in Statistics and Data Science from the Massachusetts Institute of Technology and a PhD in Systems and Computational Neuroscience from the Universidad de Guadalajara. He is currently a Research Associate at the Janelia Research Campus. Mail: nunezm@hhmi.org

§ Francisco Ramos obtained a BA in Business Creation and Development from the Monterrey Institute of Technology and a certificate in the Architecture and Engineering of Complex Systems from MIT. He is currently the Head of Data Science at Global Ecosystem Dynamics and a member of the World Economic Forum's Global Shapers Community. Mail: francisco@globalecosystemdynamics.org

¶ Olga Medrano Martín del Campo obtained a BSc in Mathematics from the Massachusetts Institute of Technology and is currently a doctoral candidate in the joint Math and CS Program at the University of Chicago. Mail: omedranomdelc@uchicago.edu

± Kevin Beuchot obtained a BSc in Mathematics from the Massachusetts Institute of Technology. He is currently a data science consultant at EY.

## 1. Introduction

Although the relationship between the term "ecosystem" and economic groups was initially a simple metaphor from biology (Moore, 1996) intended to describe the relationships between the whole and its components, recent works have advanced a conceptual framework that shows the benefits of transcending the simple analogy by highlighting that indeed, an economic ecosystem is an ecosystem in itself since it can be characterized by five components, elements, relationships, function-purpose, environment, and resources (Hoffecker, 2019; Tedesco, 2022), and can be understood and analysed as a complex, constantly evolving system (Haslanger, 2020; Arthur, 2021).

Given that it is not possible to construct effective explanatory theories using metaphors (Holland & Leinhardt, 1977), Tedesco (2022) suggests that right from the start, speaking of economic ecosystems transcends the simple analogy, and that economic ecosystems are in themselves a complex system, as described by Meadows (2008), a network created from the interacting elements, through which resources flow along a shared infrastructure, with a specific purpose or function, which, according to Odum's (1971) definition of ecosystems from the ecological sciences, are in all their characteristics one ecosystem in its entirety. It is with this in mind that Tedesco (2022) defines an economic ecosystem as "A community of actors and individuals who interact with each other and with their environment in a delimited region, which is determined by its social and natural dynamics, in which resources are exchanged with the function and/or purpose of creating some kind of economic value".

On the other hand, historically, the study of how resources flow through the network has monopolized resource-related research for decades (Benson, 1975; Granovetter, 1983; Lavie, 2008; Neumeyer & Santos, 2018; Oerlemans et al., 1998); however, it is impossible to separate economic ecosystems from the environment to which they belong and are integrated since this environment is a global ecosystem that has been developing for much longer, and in addition to including the economic ecosystem, sustains it (Tedesco & Serrano, 2019).

Paradoxically, the interactions between the elements of the economic ecosystem not only have internal consequences but also impact the global ecosystem in which they occur, as described by Haslanger (2020): "any social system are necessarily, world-involving and embodied and we have to cooperate with the world as we coordinate with each other". This interdependence has also been identified in *ecological economics* (Daly & Farley, 2011), because the very survival, development and consolidation of an economic ecosystem is contingent on the interdependence of the global, social and economic ecosystem being maintainable and sustainable (Costanza et al., 1993; Marten, 2001; Xepapadeas, 2008; Tedesco, 2022).

However, the prevailing neoclassical economic theoretical framework (Clark, 1998; Woodford, 2009), which proposes the flow of resources and competition as the main forces of accelerated economic development (Panzar, 1987; Weintraub, 2007), has resulted in highly asymmetric relationships between the elements of the economic ecosystem and an increasingly pronounced scarcity of resources (Fuchs, 2017) that directly impacts the global ecosystem (Nadeau, 2003) and therefore the survival, development, and consolidation of that very same economic system (Fox et al., 2009; Tedesco, 2020).

In contrast, when the capacities of two or more elements of the economic ecosystem are combined to solve a task or to achieve a certain objective, resources are optimized, and the growth and sustainability of the ecosystem in which these interactions occur is maintained over the long term. This type of interaction seeks the well-being of the elements of the ecosystem through collective well-being, which also has a positive impact on the global ecosystem. This type of interaction is known as *collaboration* (Kanter, 1994; Tedesco, 2022; Wood & Gray, 1991).

After a property of a complex system involved in its prosperity and development, such as collaboration, is identified, mechanisms must be found to measure this property. It is even more important to develop a metric for the ecosystemic structure of a network built from these beneficial relationships.

Note that in this study, we refer to an ecosystemic structure and/or economic ecosystem structure in the sense of any social structure, as described by Haslanger (2016): "Structures, broadly understood, are complex entities with parts whose behaviour is constrained by their relation to other parts".

Different approaches have been proposed to measure the properties of complex systems (Mata, 2020; Zinoviev, 2018). However, in the present work, this task is approached from the perspective of analysing complex networks and the properties derived from the graphs they form (we use the term network and graph interchangeably), in addition to contrasting different formulas to measure collaboration from the same graph.

This paradigm for the study of social and/or economic ecosystems not only allows the interaction of people, organizations or intangible elements of the ecosystem (such as an event that occurs within it) to be measured but also creates the possibility of using computational simulations and quantitative techniques to understand complex dynamics that would otherwise be difficult to analyse, such as the development of strongly connected communities, the growth of the ecosystem, its resilience to disruptive events, its propensity to collapse, or the general evaluation of the health of an ecosystem (Arosio et al., 2020; Huang et al., 2018).

### 1.1 Motivation for a structural metric based on collaboration

One of the main drivers of the development of a metric to measure the structures that create collaborative relationships in economic ecosystems is that collaboration is a practice that by itself can ensure that the elements of the system realize their purpose or function in it, in addition to contributing to the needs of the global ecosystem in which they are integrated.

On the other hand, from the ecological sciences perspective, there are two groups of interactions between the living (biotic) elements of an ecosystem: beneficial and nonbeneficial. In the former, all participants benefit (mutualism, cooperation), and at most only one these are harmed (commensalism). In the second group, at least one of the participants is harmed (predation, parasitism, herbivory), with competition being the most harmful of all relationships, whereby each participant is harmed in the interaction (Le Roux et al., 2020; Wootton & Emmerson, 2005).

Beneficial relationships play a critical role in ecosystem equilibrium, that is, in its homeostatic capacity, the possibility of maintaining equilibrium under changing environments or internal disruptions. The greater the number of beneficial relationships and the lower the number of nonbeneficial relationships, the more likely its homeostatic capacity is to be more robust and resilient (Ernest & Brown, 2001; Thébault & Fontaine, 2010). Therefore, there is a greater possibility that life will prosper and evolve (Cleland, 2011; Dyke & Weaver, 2013; Lovelock & Margulis, 1974).

There is no reason to think that economic ecosystems cannot function in the same way, having the same properties that we have already mentioned, of all complex systems.

In this way, since collaboration is a cooperative mechanism, measuring this, as well as its resulting structures and establishing interventions that incorporate it as an objective metric, can help strengthen the system, its connections and its performance in general.

On the other hand, it is known that the conditions of a complex system, such as resilience, robustness, efficiency and propensity to collapse constitute its structural properties and a precise way of measuring its maturity (Tedesco, 2022; Thurner et al., 2018; Westphal et al., 2007), so for this work, we propose that these same conditions as a whole, added to the average amount of collaboration in an economic ecosystem, determine its potential homeostatic capacity.

That said, this article aims to establish a way to quantify collaboration and mathematically describe the structure that is built from it in an economic ecosystem from the perspective of complex network analysis.

There are some other studies with a similar objective but with a different focus or object of study, such as the case of Westphal et al. (2007), who studied the capacity for collaboration in virtual organizations based on their level of communication, commitment, capacity response, flexibility and reliability, or the SCOR model of NC State University (SCRC SME, 2004), which establishes collaboration through performance indicators in a model of supply chain operations between partner organizations.

However, a method for evaluating collaborativeness and its underlying structures in the framework of economic ecosystems is lacking; in fact, it is not easy to even propose an analogy between the examples cited above and the characteristics of an economic ecosystem since mechanisms and properties within an organization or between partner organizations are not always of the same nature as those of the interactions within an economic ecosystem.

Additionally, we note that some of the properties proposed in these studies may have similarities with the metrics of the graphs resulting from economic systems studied from the perspective of complex systems, and which are also similar to comprehensively measuring the *health* of this type of system (Meadows, 2008). Of these properties we can highlight the effectiveness in the communication between the elements of the system, its structural robustness and its resistance to collapse (Thurner et al., 2018).

Thus, we have seen that it is valuable to combine metrics in a formula that returns a value that indicates how collaborative an economic system is and the structure that defines its homeostatic capacity as a result of the mathematical properties of its representation as a graph.

On the other hand, Tedesco (2022) not only proposes studying collaboration in economic ecosystems but also suggests a series of interesting metrics to measure it from the analysis of complex networks (graphs) and finally a compendium of suggested metrics that can utilized as indicators of the properties of the complex system (efficiency, robustness, propensity to collapse, etc.).

Having established the importance of measuring collaboration and its resulting structures, we now focus on developing the theoretical context from mathematics to achieve a single indicator that allows us to determine the structural condition of an economic ecosystem based on collaborative relationships, using various proposed formulas taking into account the properties to be measured of complex systems and their metrics of interest (Meadows, 2008; Tedesco, 2022; Thurner et al., 2018).

## 2. Origin of the data

The work presented in this document is based on data taken from a series of studies carried out with Global Ecosystem Dynamics (GED), an organization that has published a series of reports focused on practitioners, innovators, entrepreneurs and public policy-makers of economic ecosystems. This article uses the data collected by GED for the Autonomous City of Buenos Aires, Santiago de Chile, Montevideo, Sao Paulo, Mexico City, Aguascalientes, Pachuca, Oaxaca, Guadalajara, Madrid, and Valencia.

The primary data were obtained from a methodological combination that included participatory workshops with members of the economic ecosystems studied between June 2019 and February 2020 (Tedesco et al., 2020a, 2020b, 2020c, 2020d, 2020e, 2020f).

Although this article is concerned with a broader array of economic ecosystems, the data collected for this analysis belongs strictly to the economic subecosystem of innovation-based entrepreneurship in the aforementioned cities. The concept of the economic subecosystem and its relationships with the entire economic ecosystem are detailed by Tedesco and Serrano (2019), Tedesco (2020), and Tedesco (2022).

In this work, we use the term "actor" to refer to any organization or initiative of an organization that exists because of and for the benefit of the economic ecosystem to which it belongs (Hoffecker, 2019; Tedesco & Serrano, 2019). The selection of participants for the aforementioned studies was carried out through desk research, which involved a meticulous search for each economic ecosystem studied to identify the largest possible number of members of the corresponding ecosystem through different means, such as web presence, social networks, databases available from governmental and nongovernmental organizations, and local contacts, among others.

The actors participating in the GED studies are classified based on the TE-SER model (Tedesco & Serrano, 2019) to gain a higher layer of analysis, which allows the identification of the composition of an economic ecosystem based on the role that the actor potentially plays in it and the value it brings to it as a whole.

For data collection, a quantitative and qualitative instrument focused exclusively on collaborative relationships between participating actors was used, since this work focuses exclusively on collaborative relationships between actors in economic ecosystems and not on all relationships that potentially exist in these systems (Tedesco, 2022).

This instrument was used to generate networks by having participants from each ecosystem mention up to 25 of their most relevant collaborations with other actors over the last 3 years. The following figure (Figure 1) shows the visualization of the economic ecosystem of an innovation-based entrepreneurship in Valencia (Spain) through the Gephi ForceAtlas2 visualization algorithm, showing the dynamics of collaboration between the *gravitational centres* (Tedesco, 2022) in the centre of the graph.

The metrics produced by these types of graphs are used in this article.

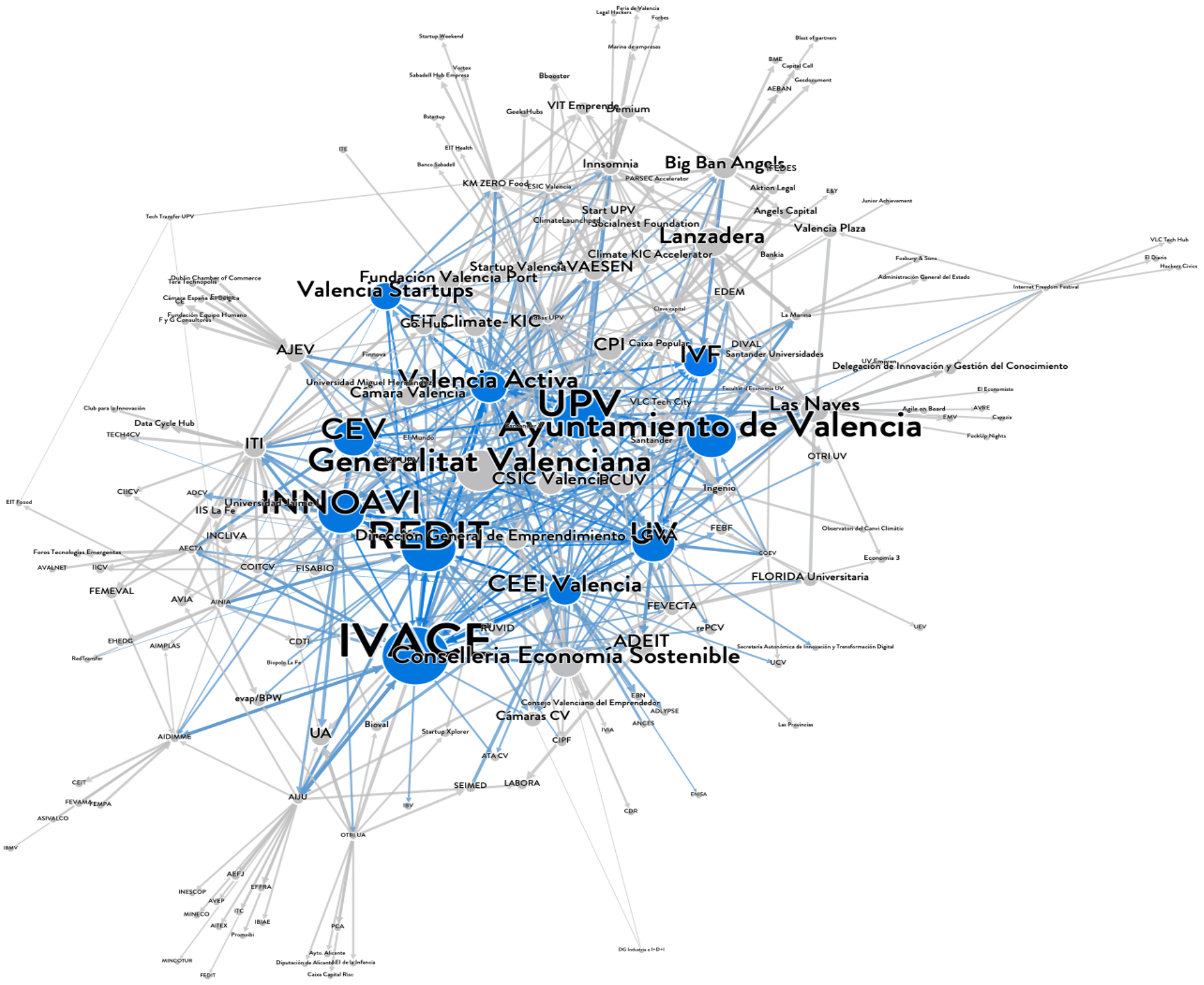

**Figure 1.** Collaboration graph from the data collected in Valencia, Spain

## 3. Theoretical framework

Before establishing the ideal properties to consider an ecosystem structural metric based on its collaboration relationships as "good", we first need to specify some definitions in terms of the mathematical language that we use:

**Definition 1.** A **graph or network** is a collection of vertices and edges that represent the relationship between elements of a set, $G_1 = (V_1, E_1)$ (Trudeau, 1993).

**Definition 1.1** a graph is **simple** if

- It is not directed: its edges have no directionality
- It is not weighted: its edges do not have weights associated with them
- It does not have parallel edges: between each pair of vertices, there is no more than one edge
- There are no loops: there are no vertices connected to themselves.
-

**Definition 2.** Given two simple graphs, these graphs are isomorphic if there is a bijective function $\varphi: V_1 \to V_2$ such that $(u, v) \in E_1$ is an edge of $G_1$, if and only if $(\varphi(u), \varphi(v)) \in E_2$ is an edge of $G_2$ for any two vertices $u, v \in V_1$.

**Definition 3.** Given a positive integer n, we define $\boldsymbol{G_n}$ as the family of all possible graphs with $n$ vertices, up to the isomorphism. This means that if two graphs $\boldsymbol{G_1}$ and $\boldsymbol{G_2}$ with $n$ vertices are isomorphic, then these are considered the same element of $\boldsymbol{G_n}$.

The following figure (Figure 2) illustrates a family $G_4$ that consists of 11 elements. Note that labelling the vertices to exclude isomorphism would produce a total of $2^6 = 64$ graphs for this family.

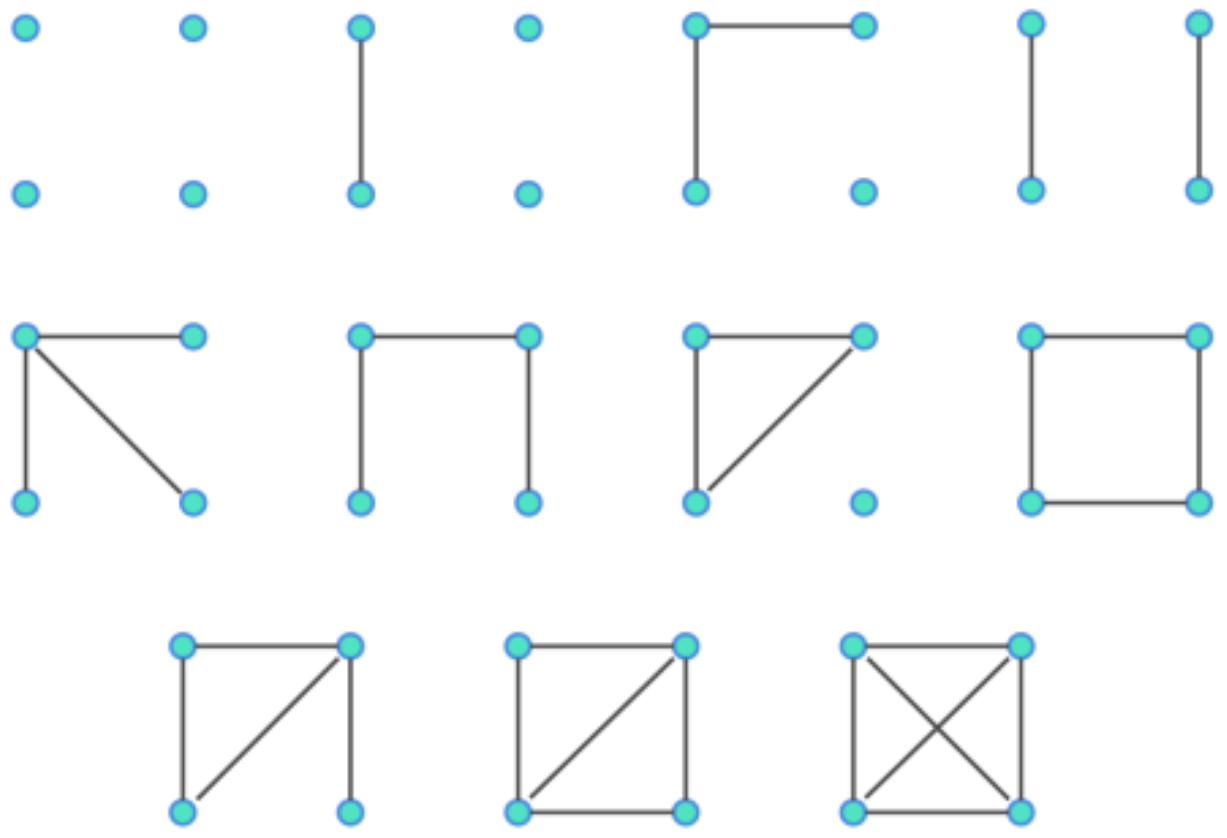

**Figure 2.** Four-node graphs in $G_4$

**Definition 4**. A **metric** is a family of functions$\{\mu_n\}_{n\in Z}$:

$\mu_n: G_n \to R$

That is, a metric is a function that maps a property of a family of graphs to the set of integers.

### *3.1 Graph metrics*

From the graphs resulting from the data collection, different metrics can be obtained that describe various properties of the data; thus, they also help us understand its structure and particularly how these metrics represent collaboration in said ecosystem.

Although we do not delve into rigorous details about each of these metrics, it seems pertinent to provide a description of what they measure and how they are interpreted in the context of economic systems.

The following table (Table 1) provides a summary of the most common metrics that we can study and measure in a network in which the interactions correspond to collaborations of the represented system. The table shows a brief description of each metric, from the perspectives of both graph theory and complex economic systems.

**Table 1.** Common CNA metrics and their usefulness in measuring economic ecosystems.

| Metric * | CNA Meaning * | Proposed interpretation for economic ecosystems[**] | Useful feature(s) to be measured | Values |
|---|---|---|---|---|
| **Average shortest path length** | The average number of steps or connections throughout the shortest routes for all possible pairs of nodes on the network. | The average number of contacts or connections that separate an actor from any other actor in the ecosystem. | Ecosystem efficiency. | From 1 to n-1 |
| **Central point dominance** | Average of the differences among the betweenness centralities metrics for all nodes to the maximum betweenness centrality in the graph. | How centralized the system is. How much power the most influential actor in the ecosystem has. | Ecosystem proneness to collapse | From 0 to 1. |
| **Clustering coefficient** | Centrality is assigned to a node via the density of its egocentric graph. The graph is restricted to its neighbours. | The extent to which collaboration is observed among the collaborators of an actor. | Ecosystem robustness | From 0 to 1 |
| **Global efficiency** | Inverse of the average characteristic path length among all nodes in the network. | How well information can travel through the network. | Ecosystem efficiency. | From 0 to 1 |
| **Average eccentricity** | Average of the centrality that assigns to each node the shortest path length possible to another node in the graph. | How far an actor can be from another within the ecosystem itself. | Ecosystem compactness and size. | From 1 to n-1 |

[**] While the mathematical meaning of the metrics remains unchanged, the final interpretation depends, to a large extent, on the type of network that is being constructed (Hanneman & Riddle, 2005).

| | | | | |
|---|---|---|---|---|
| **Average degree** | The average number of edges incidental to the total number of edges to the node; the total number of edges in the network divided by the total number of nodes in the network. | The average number of collaborations in which an actor has participated. | Number of collaborations within the ecosystem. | From 0 to infinity |
| **Modularity** | The fraction of edges that fall within the given groups minus the expected fraction, if edges are distributed randomly. | How sparse the connections are among different modules or communities in the system. | Ecosystem proneness to collapse | From -1/2 to 1 |
| **Average weight of edge** | The average weight or strength of the edge incidental to a node in the network. | The average intensity of an actor's collaborations. These intensities can depend on different factors such as time, allocated resources, and type of relationship as decided during the study. | Collaboration intensity. | From 1 to 5<br><br>*Self-defined range* |
| **Transitivity** | Fraction of all possible triangles shown in the graph; density of triangles. | How likely it is that two actors collaborating with the same actor are also in collaboration with each other. | Ecosystem robustness | From 0 to 1 |
| **Rich club coefficient** | Maximum density possible of the graph if restricted to the nodes with degree at least $k$, up to election of $k$. | How much collaboration can be observed among the most active agents of the ecosystem. | Ecosystem robustness | From 0 to 1 |

*Tedesco (2022) based on Börner et al. (2007) and Hernandez and Van Mieghem (2011)

### *3.2 Ideal properties of a structural metric based on collaboration relationships*

We propose three ideal properties that a new metric capable of measuring collaboration and its structure within an economic ecosystem from the context of the study of complex systems should have:

- **Consistency:** By this, we mean that the value of the metric is not affected by the particular structure of the system studied; in other words, if two graphs are equal even after mixing the node labels, the value of the metric must remain the same.
- **Delimitation:** it is desirable that the metric has an explicit range, that is, that the value of the function is delimited.
- **Effectiveness:** in general terms, less collaborative networks should be evaluated with a lower real value than more collaborative networks.

Obtaining the first two properties is relatively simple, in the case of consistency, since the nodes, edges and the relationship between them are always the same; even when mixed, the metrics remain consistent, in addition to the fact that the new metric is composed of in turn by metrics that are consistent, while delimitation can be achieved if the graph metrics are incorporated in the formula in a way in which each component is bounded.

Effectiveness, on the other hand, is a more complex property; first, to achieve it, we must establish what makes a graph in $G_n$ poorly collaborative and its obvious counterpart, what a totally collaborative graph would look like. In that sense, the most "natural" responses would be a graph without connections and a totally connected one.

We must take into account that a metric must always induce *complete order* in a family of graphs $G_n$ given that the target space of the metric $\mu_n$ is R, and due to the trichotomy property (Apostol, 1967), we can ensure that for any pair of graphs $G_1, G_2 \in G_n$, any of the following conditions must be met:

$\mu_n(G_1) < \mu_n(G_2), \mu_n(G_1) = \mu_n(G_2), \mu_n(G_1) < \mu_n(G_2);$

However, there may be cases for which it is difficult to determine which of the two graphs is more collaborative based only on the number of edges (Figure 3).

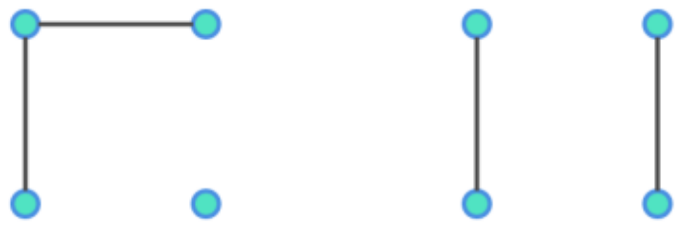

**Figure 3.** Which of the two graphs is more collaborative?

The previous observation raises the need to develop a methodology that allows us to know if a collaboration metric is effective or not. In the present work, we chose to use computationally generated synthetic graphs to test the proposed collaboration metrics.

## 4. Methodology

### 4.1 Construction of the synthetic graphs

As indicated in the previous section, the experimental instrument used to test the effectiveness of each of the possible formulas proposed to serve as a collaborative metric is a collection of 800 computationally generated synthetic graphs, each with between 120 to 400 nodes[††].

Each of these graphs is constructed by simulating the data collection performed in Tedesco et al. (2020a, 2020b, 2020c, 2020d, 2020e), including the nodes in the graph and their respective connections. Each node represents an organization in the research instrument (survey), with a probability of being surveyed of *prob_resp*, as long as the number of respondents remains within *respondents_range* and that in turn reports a number of collaborations (edges in the graph) within the *connections_range.*

Each of these collaborations connects the organization (node) in question with a new one with a probability *prob_new.* A node (actor) that does not respond to the survey reports with probability *p = 0.25,* from 1 to 6 edges with nodes that already exist in the graph.

The set of 800 graphs can be subdivided into four sets of 200, each of which is generated based on the variation of exactly one of the four parameters introduced below and fixing the other three.

[††] https://github.com/OlgaMMC/iGED_new_repo

**New connections**: The probability that an edge n goes to a new node (*prob_new*) in the construction of graph $G_k, 1 \leq k \leq 200$ varies linearly from .70 to .30 as follows:

- $0.55 + \frac{100-k}{400}, 1 \leq k \leq 100$
- $0.55 - \frac{k-100}{400}, 101 \leq k \leq 200$

The motivation for introducing this variability is based on the fact that the less likely it is that a respondent mentions a collaboration with an agent recently mentioned in the system, the more likely the resulting graph is denser, which results in a more robust structure and, therefore, greater collaborative activity. Therefore, the higher the k index is, the more collaborative we expect our ecosystem to be.

**Number of responses**: The range of the valid number of collaborations (*connections_range*) that a node can report in the construction of $G_k, 1 \leq k \leq 200$ varies as follows:

- $[15 - \frac{100-k}{10}, 24 - \frac{100-k}{10}]\ 1 \leq k \leq 100$
- $[15 + \frac{k-100}{10}, 24 + \frac{k-100}{10}]\ 101 \leq k \leq 200$

The motivation for this variability is based on the fact that the greater the number of collaborations reported by each node surveyed, the greater the connectivity that we expect from the resulting graph. Therefore, when an index k is higher, we expect our graph to be more collaborative.

**Range of respondents**: The total possible number of respondents (*respondent_range)* during the construction of the graph $G_k, 1 \leq k \leq 200$ varies as follows:

- $35 - \frac{100-k}{4}\ 1 \leq k \leq 100$
- $35 + \frac{k-100}{4}\ 101 \leq k \leq 200$

This variability is motivated by the fact that the greater the number of nodes surveyed, the greater the amount of information collected and, in particular, the greater the number of connections. Therefore, as k increases, we expect our graph to become more collaborative.

**Respondents**: The probability that a node added to the graph is a respondent varies as follows:

- $.04 - \frac{100-k}{400}\ 1 \leq k \leq 100$
- $.04 + \frac{k-100}{400}\ 101 \leq k \leq 200$

This variability is motivated by a reasoning similar to the previous one: the greater the possibilities that we have that a node responds to the survey, the more connections we see in the resulting graph. Therefore, the higher the k index is, the more collaborative we expect our graph to be.

In the following section, we describe a set of 15 proposed formulas to measure the collaborative structure of an economic ecosystem from the graph that represents it, in addition to detailing the way in which we measure its performance in terms of its effectiveness with the use of synthetic graphs detailed here.

### 4.2 Proposed metrics

Thinking about representative metrics and proposing new formulas for their measurement is a learning process since each formula has its own level of complexity and tries to reflect the *quantity* and *quality* of collaboration given the structure of the graph that describes the structure of the economic ecosystem, taking into account small variations.

Originally, a formula ($C_0$) was proposed as a basis that included the metrics of *average number of collaborations (avg.colabs), clustering (clus)* and *modularity (mod)* since these metrics satisfactorily describe properties of a complex system such as resilience and propensity to collapse and the efficiency of the flow of interactions between the various nodes, among others. The discussion about why combining these graph metrics can lead to a good metric for the structure of an economic ecosystem is detailed in Tedesco (2022).

$$C_0\,(G) := \ avg.colabs(G) \cdot (clus(G) - log_{10}(mod(G)^2))$$

One of the main modifications to all the proposed formulas as a result of $C_0$ is the inclusion of the *global efficiency (efi.global)* as one of its components since this graph metric provides a specific idea of how efficient communication is in the network is, while at the same time describes the efficiency property proposed for any complex system. Mathematically, the global efficiency is an average of all the global node efficiencies, each of which is, in turn, the average of the reciprocal of the shortest path lengths to all other nodes.

That is, this metric positively describes the communication in the network and potentially how efficient the transfer of resources between the nodes (actors) is: *the shorter the geodesics (minimum length line that joins two nodes within the graph),* the higher the reciprocal values are, and therefore, the greater the overall efficiency of the network is.

$C_1$ is defined as:

$$C_1(G) := \ efi.global(G) \ + \ trans(G) \ + \ (1 - \frac{mod(G) + core(G)}{2})$$

The second component of this formula is given by *transitivity* (*trans)*, which positively describes the robustness of the network. It describes to some extent the adaptability of the network to changes, such as the elimination of some nodes and edges, which may impact the future resilience capacity and its homeostatic equilibrium.

The last component of this formula is determined by the *modularity* and the *core ratio (core).* These terms are averaged and subtracted from 1 because each of the metrics negatively describes the collaborative network.

The lower the value of modularity is, the lower the segregation and the greater the incentive for agents in different regions of the system to collaborate with each other. In addition, the higher the *core ratio* is, the greater the level of reach in the network, which allows more collaborations to occur in the future.

In this sense, the formula $C_2$ differs from $C_1$ only in the third component, the difference aims to describe the preparation of the network for future developments. On the one hand, if the reciprocal of the *eccentricity (exc)* is greater, it implies smaller distances between pairs of nodes, which facilitates the emergence of new collaborations by proximity; on the other hand, the modularity is linearly integrated with a coefficient of -1. The lower the modularity in a graph, the less segregation is observed, which makes it more likely that this network has collaborations between the different regions making up the graph.

$$C_2(G) := \ efi.global(G) \ + \ trans(G) \ + \ \frac{1}{exc(G)} - mod(G)$$

In contrast, in $C_3$, the communication component is still described by *global efficiency*, but we use *clustering* to approximate robustness and the propensity to collapse is evaluated through *modularity.* $C_3$ is then comparable to $C_1$ in that average collaborations are substituted for global efficiency, modularity is included, and the components are geometrically averaged.

$$C_3(G) := \ \sqrt[3]{efi.global(G) \ \cdot \ clust(G) \ \cdot \ \frac{1}{2} \ \cdot \ (1 \ + \ cos(\pi \cdot mod(G)))}$$

This formula emulates some of the properties of the $log_{10}(x^2)$ function used to include modularity in $C_0$, including a monotonic increase over [-1,0] and a monotonic decrease over [0,1], with an important difference, that is, the function

$$f(x) = \frac{1}{2} \cdot (1 + cos(\pi \cdot mod(G))$$

is normalized such that $f(x)$ is included in [0,1] for all $x \in [0,1]$, unlike the logarithmic function, which is infinite.

The formula $C_4$ is close to the previous one; differing only in that the robustness component is given by transitivity instead of clustering. We test this configuration because transitivity turns out to be a better metric for the purposes of this study since the *clustering* metric suffers from the problem that all values per node are calculated relatively; that is, the lower the degree of a node, the more likely it is that its neighbourhood is populated and the greater the clustering (Zinoviev, 2018). This limits the robustness approximation to terms of the degree of the node, while we are also interested in understanding whether the role that this node plays in the structure of the graph is relevant.

$$C_4(G) := \sqrt[3]{efi.global(G) \cdot trans(G) \cdot \frac{1}{2} \cdot (1 + cos(\pi \cdot mod(G)))}$$

In $C_5$, we decide to try for the first time not to incorporate modularity; the components of communication and robustness are given by global efficiency and transitivity, respectively; however, in this proposal, we use the *core ratio* as a property that contributes to the ecosystem's propensity to collapse. The lower the value is, the greater the potential for future collaborations in the network.

$$C_5(G) := efi.global(G) + trans(G) + (1 - core(G))$$

The formula $C_6$ is very similar to $C_4$, with the difference being that the propensity to collapse is evaluated by means of the eccentricity (*exc).*

$$C_6(G) := \sqrt[3]{efi.global(G) \cdot trans(G) \cdot sin(\frac{\pi}{exc(G)})}$$

Eccentricity is added this way because, as an integer, $\frac{\pi}{exc(G)} \in [0, \pi]$, the sine of this number must be in the range [0,1].

From this formula, the average number of collaborations is one of the metrics considered for taking the study framework and the measurement instruments into account since in the series of studies by Tedesco et al. (2020a, 2020b, 2020c, 2020d, 2020e, 2020f), the information was obtained through the snowball sampling method (Zinoviev, 2018), which has a notorious implication: it is possible that the observed data do not reflect the totality of what is happening in the real world system.

This is a common challenge that arises in the modelling of any complex system or even only in the study of any real world system (Kossinets, 2006). The decision to measure the amount of collaboration in these networks based on the average collaborations reported through data collection per se aims for transparency regarding the amount of collaboration observed by the measurement instrument.

$C_7$ is defined as follows:

$$C_7(G) := \frac{1}{2}(ln(1 + \frac{avg.collabs(G)}{m}) + \frac{efi.global(G) + trans(G) + \frac{1}{exc(G)}}{3})$$

Let us recall that *m* represents the maximum number of collaborations that a single node can report in the system. This formula averages only two components: quantity of collaboration and quality or structure of collaborations.

The quantity is determined only by the average number of collaborations, while the quality considers three subcomponents. Similar to the previous formulas, the overall efficiency describes the communication, the transitivity describes the robustness, and the reciprocal of the eccentricity describes the networks resistance to collapse.

The only difference between $C_8$ and the previous index regards the collapse resistance property, since in this formula, it is measured through *modularity*, which is incorporated through a cosine function, in the same way as in the formula $C_3$.

$$C_8(G) := \frac{1}{2}(ln\,(1 + \frac{avg.collabs(G)}{m}) + \frac{efi.global(G) + trans(G) + \frac{1}{2} \cdot cos(\pi \cdot mod(G))}{3})$$

The composition of formula $C_9$ is close to that of $C_7$, except that instead of considering the quantity and quality of collaborations as the two main components, it considers the quantity of collaborations at the same weight as the communication, robustness, and resistance to collapse of the graph.

$$C_9(G) := \frac{1}{4}(ln\,(1 + \frac{avg.collabs(G)}{m}) + efi.global(G) + trans(G) + \frac{1}{exc(G)})$$

The formula $C_{10}$ logarithmically incorporates the average of collaborations obtained in the data collection process, like all formulas starting from 7, and geometrically averages the three subcomponents of collaboration quality, in the same way as in $C_6$.

$$C_{10}(G) := \frac{1}{2}(ln\,(1 + \frac{avg.collabs(G)}{m}) + \sqrt[3]{efi.global(G) \cdot trans(G) \cdot sin(\frac{\pi}{exc(G)})})$$

From $C_{11}$, the average number of collaborations is considered equally important as the other components that describe the quality or structure of the collaboration. In addition, the specific case of $C_{11}$ incorporates the average of collaborations as the square root of its fraction over the maximum possible number of reportable collaborators (*m*), in addition to incorporating the *rich club coefficient* and the *core ratio,* in addition to the reciprocal of the eccentricity to describe the system's propensity to collapse*:*

$$C_{11}(G) := \frac{1}{4}(\sqrt{\frac{avg.collabs(G)}{m}} + efi.global(G) + trans(G) + \frac{r.c.c.(G) \cdot core(G)}{exc(G)})$$

In the last three formulas, we begin to consider the metric called *central point dominance* (cpd). In the case of the formula $C_{12}$ The logarithmic average of collaborations is included, which represents the amount of collaboration; communication is represented by global efficiency; and robustness is represented by the following geometric average:

$$\sqrt{trans(G) \cdot (1 - c.p.d.(G))}$$

In this sense, robustness is captured in terms of density of triangles, and its resilience is obtained by subtracting the *central point dominance* from 1*,* which describes the dependence of the graph relative to its most influential node. Finally, the propensity to collapse is given by the reciprocal of the eccentricity:

$$C_{12}(G) := \frac{1}{4}(ln(1 + \frac{avg.collabs(G)}{m}) + efi.global(G) + \sqrt{trans(G) \cdot (1 - c.p.d.(G))} + \frac{1}{exc(G)}$$

In $C_{13}$, we add a small variation where instead of including the average of collaborations logarithmically, it is included through the square root of the average over the maximum possible of reportable collaborators.

$$C_{13}(G) := \frac{1}{4}(\sqrt{\frac{avg.collabs(G)}{m}} + efi.global(G) + \sqrt{trans(G) \cdot (1 - c.p.d.(G))} + \frac{1}{exc(G)})$$

Finally, $C_{14}$ is the most complex formula of all, considering 7 variables, and describes the amount of collaboration by means of the square root of the average collaboration between the maximum possible reportable collaborators, communication as the square root of the overall efficiency between the eccentricity, the robustness by the geometric average used in $C_{12}$, and the propensity to collapse given by the product of the *rich club coefficient* and the *core ratio.*

$$C_{14}(G) := \frac{1}{4}(\sqrt{\frac{avg.collabs(G)}{m}} + \sqrt{\frac{efi.global(G)}{exc(G)}} + \sqrt{trans(G) \cdot (1 - c.p.d.(G))} + (r.c.c(G) \cdot core(G))$$

To summarize all the proposed formulas, we generate a diagram that taxonomically arranges them (Figure 4).

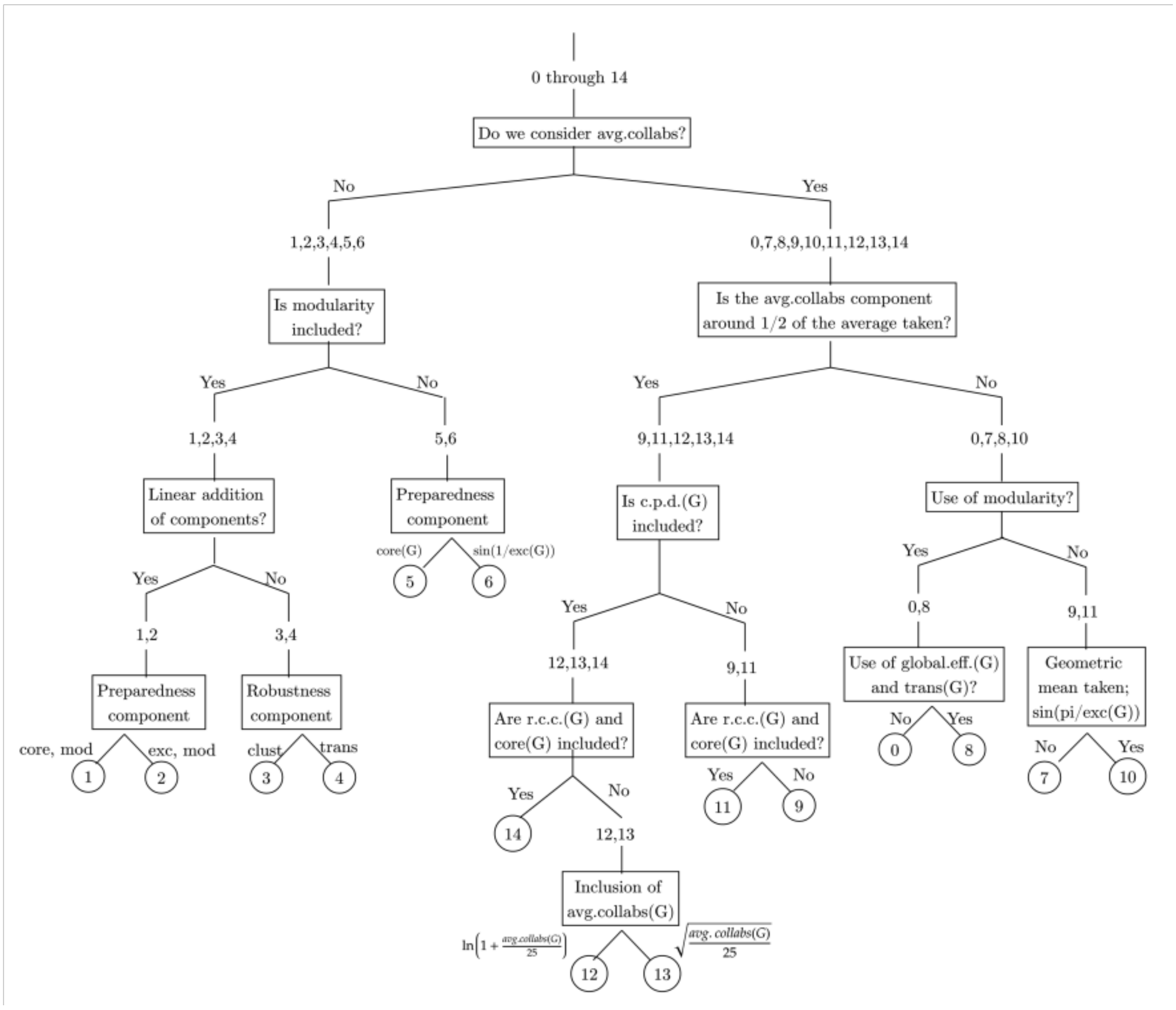

**Figure 4.** Taxonomic diagram of the proposed formulas based on the metrics that compose them.

### 4.3 Evaluation of the formulas with the synthetic graphs

Now that we have presented how the synthetic graphs are constructed and the formulas to be evaluated, we can focus on the evaluation process. All the proposed formulas are consistent; since they are composed of metrics that are consistent, they are also delimited since the components are added to the formulas in a normalized way so that the effectiveness is the remaining property to be evaluated.

To carry out each collaboration, formula $C_i$ is evaluated under the following procedure for each of the four sets of synthetic graphs mentioned in Section 4.1:

1. The 200 corresponding graphs are numbered $G_k, k = 1,2,...,200$.
2. The collaborative formula $C_i(G_k)$ is evaluated for each$G_k, k = 1,2,...,200$.
3. The graphs are classified based on the values obtained in the previous step from lowest to highest.
4. Ideally, if the variability in the construction of the synthetic graph is positively correlated with the collaborativeness measured by the proposed formula, a good structural metric of collaboration, in this case, should have effectiveness indices of at least 0.95, which we call *good performance*.

## 5. Results

Each of the proposed formulas perform differently based on the test procedures described above, with some formulas being more effective than others (Table 2).

**Table 2.** Effectiveness of each of the structural collaboration formulas for each of the construction variabilities in the synthetic graphs.

| Metric | New connections | Number of responses | Range of respondents | Respondents |
|---|---|---|---|---|
| Original (0) | 0.975 | 0.8 | 0.375 | 0.55 |
| 1 | 0.975 | 0.375 | 0.325 | 0.875 |
| 2 | 1 | 0.325 | 0.3 | 0.825 |
| 3 | 1 | 0.4 | 0.35 | 0.575 |
| 4 | 1 | 0.275 | 0.4 | 0.675 |
| 5 | 0.7 | 0.525 | 0.375 | 0.925 |
| 6 | 1 | 0.375 | 0.25 | 0.9 |
| 7 | 0.95 | 0.85 | 0.325 | 0.675 |
| 8 | 0.95 | 0.85 | 0.35 | 0.625 |
| 9 | 1 | 0.8 | 0.35 | 0.8 |
| 10 | 1 | 0.825 | 0.325 | 0.875 |
| 11 | 1 | 0.85 | 0.325 | 0.8 |
| 12 | 1 | 0.8 | 0.3 | 0.8 |
| 13 | 1 | 0.8 | 0.3 | 0.8 |
| 14 | 1 | 0.825 | 0.325 | 0.825 |

Based on the previous table, the formulas with the best performance are Formulas 10 and 14, and Formula 10 not only performs better in the subset of respondents but also uses fewer metrics than Formula 14.

Considering this process, the proposition of Formula 10 supposes an improvement with respect to the original formula that focuses on the amount of collaboration and not on the structure and quality of it, the latter being what Tedesco (2022) proposed as relevant for understanding the social dynamics of an economic ecosystem and its homeostatic capacity. Similarly, it measures collaborativeness not only in terms of the new connections and the number of responses but also taking into account the data collection process, having better effectiveness in terms of the probability that a node added in the graph is also respondent (*snowball sampling*).

On the other hand, the number of respondents (*respondent range),* which represents the nodes added in the graph that are also surveyed later, seems not to be relevant to evaluating collaboration since no formula was related to that subset. This means that the number of nodes (actors) that respond to the research instrument (survey) does not seem to be relevant in terms of obtaining the structural description of the ecosystem from the theory of complex systems and the use of complex network analysis.

By taking Formula 10 as the best for measuring the structure built by collaborating, we set ourselves the task of rescaling it in the range of [1,10]. To do this, we establish the theoretical limits of the formula. On the one hand, the average of collaborations is between 0 and *m,* so the corresponding component in the formula has a value that goes from [0, 0.347].

$$\frac{1}{2} \cdot ln(1 + \frac{0}{m}) \ = \ 0; \frac{1}{2} \cdot ln(1 + \frac{m}{m}) \ = \ \frac{ln(2)}{2} \simeq 0.347$$

On the other hand, both the global efficiency and the transitivity of a graph are in the range of [0,1] and $\frac{\pi}{exc(G)} \in [0, \pi]$, which implies that $sin(\frac{\pi}{exc(G)})$ is also found in [0,1]; therefore, the second component of the formula, which is half of a geometric average of three values between 0 and 1, is between [0,0.5], so we can conclude that for any simple nondirectional graph:

$$C_{10}(G) \in [0, \frac{ln(2) + 1}{2}] \simeq [0,0.847]$$

Only with linear rescaling can we obtain the desirable limits [1,10]: $C_{10,r}(G) := \ 1 \ + \ 10 \cdot C_{10}(G)$.

We can see the collaborative structural measurements by the formula, both rescaled and not rescaled for the most collaborative ecosystem (Valencia, Spain) and least collaborative ecosystem (Sao Paulo, Brazil) sampled until the moment of being analysed by the Global Ecosystems Dynamics Initiative (Tedesco et al., 2020a, 2020b, 2020c, 2020d, 2020e, 2020f).

The above is consistent with internal team evaluations before this work using the original formula and highlights the importance of having a formal way of measuring this property of economic ecosystems - or to study the cooperative/collaborative structure of other types of ecosystems - because “naturally”, when viewing ecosystems, there may be a "natural" and subjective tendency to say that one is more collaborative than another or that one has a better ecosystemic structure than the other, and being able to quantify it opens a broader panorama of questions, and, more importantly, a way to find answers (Figure 5).

An example of this is the case of Mexico City, which although it has a lower density than Sao Paulo, that is, it has fewer connections in total between all the actors that are part of the network, its structural index suggests superior collaboration, which achieves the objective of having a formula that allows us to evaluate an ecosystem taking into account not only the number of connections but also the presence of a structure that facilitates the exchange of information and resources between the different actors of the network, which can finally provide economic ecosystems capable of maintaining equilibrium while allowing its members to prosper.

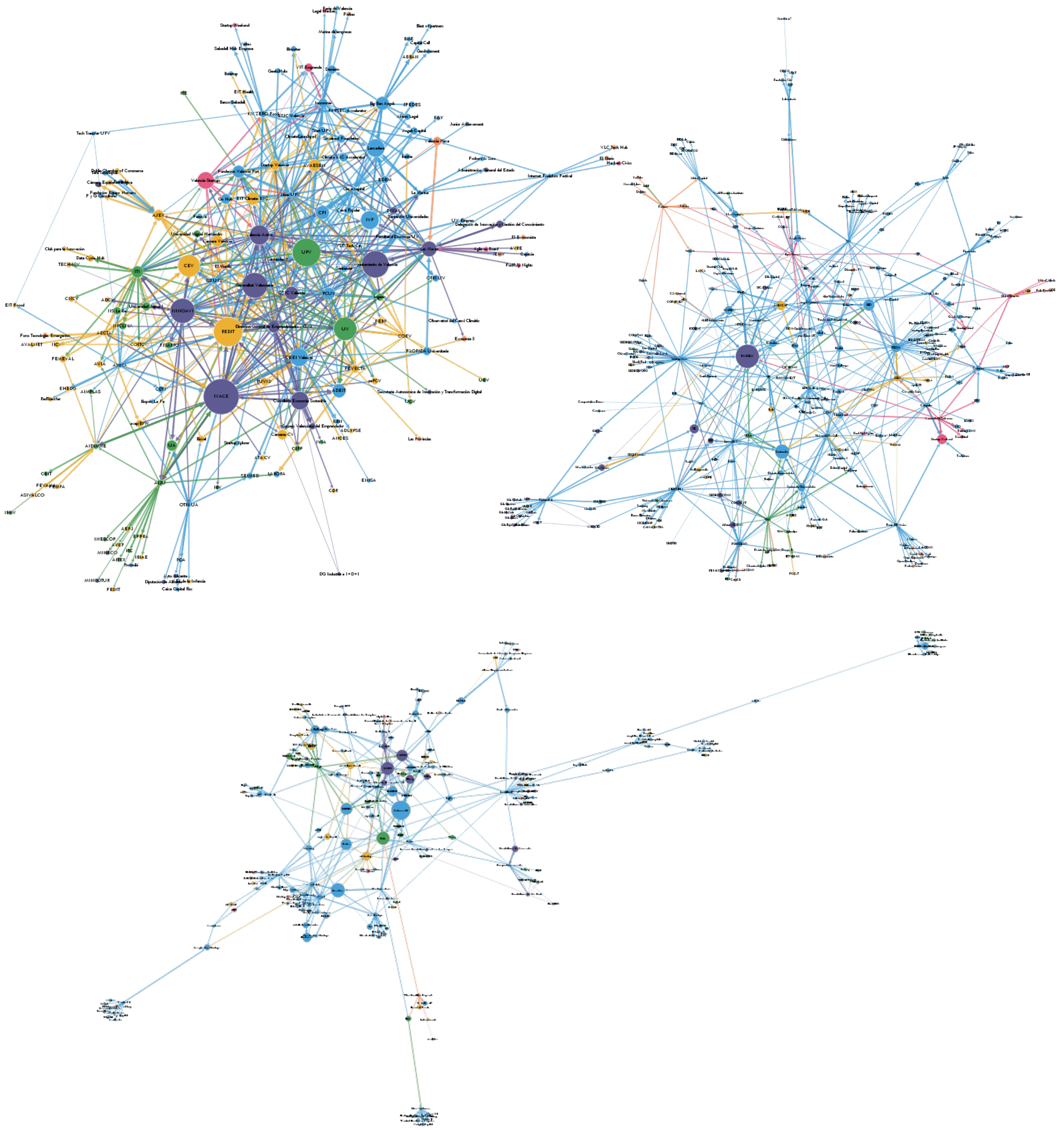

**Figure 5.** Economic ecosystems of Valencia, Spain (upper left), with its collaboration metrics given by $C_0 = 9.32, C_{10} = 0.63$ $and$ $C_{10,r} = 7.31$; Mexico City, Mexico (upper right), with its collaboration metrics given by $C_0 = 3.64, C_{10} = 0.40$ $and$ $C_{10,r} = 4.98$; and Sao Paulo, Brazil (bottom), with its collaboration metrics given by $C_0 = 3.25, C_{10} = 0.38$ $and$ $C_{10,r} = 4.85$ respectively.

**Table 3.** CNA metrics and Collaboration Index for the innovation-driven entrepreneurial ecosystems of Sao Paulo, Mexico City and Valencia

| Metric | Sao Paulo | Mexico City | Valencia |
|---|---|---|---|
| Nodes | 216 | 299 | 180 |
| Edges | 360 | 542 | 623 |
| Shortest Average Path | 4.32 | 3.82 | 3.01 |
| Central Point Dominance | 0.24 | 0.20 | 0.12 |
| Clustering Coefficient | 0.3 | 0.18 | 0.38 |
| Density | 0.015 | 0.012 | 0.037 |
| Global Efficiency | 0.27 | 0.29 | 0.37 |
| Eccentricity | 6.73 | 5.62 | 4.23 |
| Average Degree | 3.37 | 3.66 | 6.99 |
| Modularity | 0.68 | 0.62 | 0.37 |
| Average Weight of Edges | 3.43 | 3.48 | 3.54 |
| Transitivity | 0.08 | 0.05 | 0.25 |
| Rich Club Coefficient | 0.25 | 0.20 | 0.56 |
| Average Collaborations Per Participant | 10.38 | 12.33 | 15.05 |
| Collaboration Index $C_0$ | 3.25 | 3.64 | 9.32 |
| Collaboration Index $C_{10}$ | 0.38 | 0.40 | 0.63 |
| Collaboration Index $C_{10,r}$ | 4.85 | 4.98 | 7.31 |

## 6. Conclusions and future work

Measuring aspects of complex systems, and in particular economic ones, is a challenging task but potentially provides useful information for understanding the phenomena that occur in these systems.

The analysed formulas, and in particular Formula 10, showed good performance and, therefore, can be used to describe and study the structure of economic ecosystems - or even perhaps other types of ecosystems.

The proposed formula is designed in such a way that it considers the properties of complex systems; therefore, it measures collaboration both from the quantity itself present in an ecosystem, as well as the structure that these relationships build. That is, it is capable of approximating the homeostatic capacity of the economic ecosystem based on the characteristics described by Tedesco (2022).

Given that the previous characteristics are taken from both the inherent properties of any complex system and that the dynamics of cooperation are also typical of biological ecosystems, the door opens the possibility that this formula can be tested and used in this field.

Measuring collaborativeness and structures in an economic ecosystem through its graph can help to better understand the value of beneficial interactions (Le Roux et al., 2020; Wootton & Emmerson, 2005), contributing not only to each of the organizations but also to the societies in which these ecosystems are immersed and to which they add value.

Even though all the proposed formulas were designed to work with unbounded networks with characteristics akin to the ones provided by GED, we hope these formulas can be further tested on similar networks, tailored for applicability to other kinds of economic ecosystem networks, and inspire the development of similar comparative metrics that further the study of collaboration and structures for economic ecosystems from a CNA perspective. The structure and homeostatic capacity of an economic ecosystem - or any other ecosystem – is a reflection of its health and prosperity.